\documentclass[a4paper]{article}
\pdfoutput=1
\usepackage{INTERSPEECH2020}
\usepackage{multirow}
\usepackage{diagbox}
\usepackage{float}

\title{InSE-NET: A Perceptually Coded Audio Quality Model based on CNN}
\name{Guanxin Jiang$^{1,2}$, Arijit Biswas$^1$, Christian Bergler$^2$, Andreas Maier$^2$}
\address{
  $^1$Dolby Germany GmbH, Nuremberg, Germany\\
  $^2$Pattern Recognition Lab, FAU Erlangen-Nuremberg, Erlangen, Germany}
\email{guanxin.jiang@dolby.com, arijit.biswas@dolby.com, christian.bergler@fau.de, andreas.maier@fau.de}

\begin{document}

\maketitle
\begin{abstract}
Automatic coded audio quality assessment is an important task whose progress is hampered by the scarcity of human annotations, poor generalization to unseen codecs, bitrates, content-types, and a lack of flexibility of existing approaches. One of the typical human-perception-related metrics, ViSQOL v3 (ViV3), has been proven to provide a high correlation to the quality scores rated by humans. In this study, we take steps to tackle problems of predicting coded audio quality by completely utilizing programmatically generated data that is informed with expert domain knowledge. We propose a learnable neural network, entitled InSE-NET, with a backbone of Inception and Squeeze-and-Excitation modules to assess the perceived quality of coded audio at a 48\,kHz sample rate. We demonstrate that synthetic data augmentation is capable of enhancing the prediction. Our proposed method is intrusive, i.e. it requires Gammatone spectrograms of unencoded reference signals. Besides a comparable performance to ViV3, our approach provides a more robust prediction towards higher bitrates.

\end{abstract}

\section{Introduction}

Audio quality perceived by listeners is a core performance metric in many multimedia networks, such as Voice over Internet Protocol (VoIP), Digital Audio Broadcasting (DAB) systems, and streaming services \cite{Campbell2009AudioQA}. Due to the lossy compression, the audio quality could be potentially noticeably inferior and deteriorate the end-user experience. Subjective listening tests are conducted to accurately reflect the audio quality that listeners would perceive. Objective audio quality assessment methods such as Perceptual Evaluation of Audio Quality (PEAQ) \cite{peaq} and the Virtual Speech Quality Objective Listener (ViSQOL) %\cite{Hines:ViSQOL01} 
are utilized to refrain from the heavy workload of subjective listening tests. Despite their effectiveness, such algorithms relying on traditional signal processing units are always questioned about their consistency with human judgment. Even though several objective quality evaluation systems, to some extent, are perceptually-motivated by design, they generally focus narrowly on specific speech or audio codecs and would become obsolete with the emergence of new scenarios. A more recent alternative is provided by deep-learning-based concepts, which have offered solutions that are accurate, rapidly re-trainable, and easily expandable in many speech and audio-related tasks. However, most of the present research contributions in deep learning are committed to predicting the reference-free speech quality at lower sampling rates \cite{7760662,Fu2018QualityNetAE,Ooster2018,8401691, 8683175,  NISQA, Gamper, Wawenets, MittagSiamese, Mittag2020DNNNP}. Very few research works exploit general audio signals at higher sampling rates, e.g., 44.1 or 48\,kHz  \cite{Ebrahimi2018PredictingAA, Roberts2020DeepLS}. Reference-free metrics support many applications but do not apply in cases that require comparison to the ground truth, e.g. for learned speech enhancement \cite{QualityNet}. Furthermore, such learning-based systems require a considerable amount of human-annotated data and therefore constantly suffer from a lack of annotation and poor generalization on out-of-sample data.

To address the above challenges we present Inception Squeeze-and-Excitation-Net (InSE-NET), which is, to the best of our knowledge, the first deep learning-based coded audio quality assessment model handling general audio at 48\,kHz. Convolutional Neural Network (CNN) architectures, e.g. Inception \cite{Incept1, Incept2} and ResNet-50 \cite{ResNet} are commonly used for image classification tasks due to its comprehensive feature extraction ability and therefore were adapted to large-scale audio classification tasks \cite{Sevilla2017AudioBC,HersheyIncept}. In this work, we modified the original Inception-v3 \cite{Incept2} network for our regression task, which takes advantage of the parallel structure of the Inception module. The extra Squeeze-and-Excitation (SE) \cite{SE} block is applied channel-wise to improve the robustness of the model against certain out-of-sample data with minimal complexity overhead. To the best of our knowledge, we are not aware of the usage of the Inception network for a regression task for speech and audio quality prediction. The proposed model utilizes the augmented synthetic data to improve the quality prediction towards higher bitrates. InSE-NET is an intrusive method implying that it deals with the Gammatone spectrograms of both the unencoded and the coded signals. Generally speaking, intrusive methods have higher accuracy than non-intrusive methods due to calculating the correlation between the reference and the coded signals.

In this study, we attempt to approximate ViSQOL v3 (ViV3) \cite{v3} with InSE-NET. ViSQOL \cite{Hines20155} is a speech quality evaluation model that was later adapted to audio quality evaluation (ViSQOLAudio) \cite{7940042}. ViSQOL takes in both the coded signals and their corresponding original unencoded reference and predicts Mean Opinion Score-Listening Quality Objective (MOS-LQO) scores. It has been reported in \cite{Fraunhofer}, that out of all objective measures (designed to evaluate codecs), ViSQOLAudio (ViA) shows the best correlation with subjective scores and achieves high and stable performance for all content types. ViV3 is a combined release of ViSQOL and ViA. Thus, our main motivation to approximate a state-of-the-art objective measure such as ViV3 with a Deep Neural Network (DNN) is to build a learnable coded audio quality metric without launching a large-scale listening effort to gather training data. The goal of this paper is to demonstrate that after successfully mimicking a state-of-the-art objective audio quality metric with a DNN, it is possible to improve over it even with carefully designed synthetic data. The general idea of approximating an objective speech quality metric (e.g. PESQ) with a DNN \cite{Fu2018QualityNetAE} and using it as a learned loss function for speech enhancement \cite{QualityNet} is not new. However, we are not aware of any work on approximating an objective quality of general 48\,kHz-coded audio quality metric with a DNN. We demonstrate that our novel model is not only able to accurately mimic ViV3, but also when fine-tuned with synthetic data mitigates the weaknesses of ViV3.

The most obvious application of our work is a stand-alone coded audio quality metric, which is useful e.g., for validating encoder improvements with new tunings. Furthermore, our proposed model can be developed into a learnable comparable loss function for coded audio enhancement \cite{DCAE}. The output layers of InSE-NET can also be easily adapted and retrained for new applications, such as the discriminator in Generative Adversarial Networks for coded audio enhancement \cite{DCAE} and generation tasks.

The rest of the paper is organized as follows: In Section 2, the data used for model training and evaluation are introduced. Section 3 depicts the details of our proposed model and experiment setup. The related experimental results and analysis are given in Section 4, and finally, the conclusion is drawn in Section 5. 

\section{Datasets}

\subsection{Training set}
\label{subsection:trainingset}
We use a corpus of 10\,h and 2\,h of unencoded mono music (covering 10 genres, 1\,h each) and speech excerpts, respectively, at a 48\,kHz sample rate. The unencoded reference is first segmented into 5400 (7.2\,s long) excerpts. These excerpts are then encoded and decoded by High-Efficiency Advanced Audio Coding (HE-AAC) \cite{HE-AAC} and Advanced Audio Coding (AAC) \cite{AAC} codec with a wide range of bitrates. Bitrates 16, 20, 24, 32, 40, and 48 kbps were encoded with HE-AAC, and bitrates 80, 96, and 128 kbps were encoded with AAC. We chose both AAC and HE-AAC as the codecs in the training so that both waveform coding (AAC) and parametric coding tools (e.g., Spectral Band Replication in HE-AAC) are covered. We took special care in making sure that the bitrates are well-tuned so that quality increases with bitrates. This helps in making sure that the model is capable of ranking correctly; i.e. if a signal $x_j$ is a programmatically degraded version of the same unencoded signal $x_i$, then their scores should reflect such relation ($MOS_i \geq MOS_j$) \cite{SESQA}. We also included the (3.5\,kHz and 7\,kHz) low-pass filtered versions of the unencoded signals because typically these two anchors are used in a Multiple Stimuli with Hidden Reference and Anchor (MUSHRA) \cite{MUSHRA} listening test. In total 59,400 coded signals are generated from 5400 unencoded excerpts. The uncoded reference and coded signals are aligned and paired and later fed into ViV3 to produce MOS-LQO as the corresponding ground truth labels instead of human-annotated MOS scores (a number between 1 to 5, where a higher score implies better quality).

% ABISW: figure not required as the description is clear. 
%\begin{figure}[ht]
%  \centering
%  \includegraphics[width=.75\linewidth]{IS2020_paper_kit/LaTeX/Image/Data_generation.pdf} 
%  \caption{Data generation pipeline}
%  \label{fig:data}
%\end{figure}

Gammatone spectrograms of reference and coded signals are simultaneously extracted by the front-end of ViV3. Gammatone filters are a popular approximation to the filtering performed by the ear. Gammatone-based spectrogram can thus be considered as a more perceptually-motivated representation than the traditional spectrogram. The Gammatone spectrogram of the audio signal is calculated with a window size of 80\,ms, hop size of 20\,ms, and 32 frequency bands ranging from 50\,Hz up to 24\,kHz. The resulting Gammatone spectrograms of 7.2\,s of reference and coded signals are paired and stacked along channel dimension, which results in an input size of 2$\times$32$\times$360 (channels$\times$bands$\times$time-frame) to the neural network. While one may consider other perceptually motivated spectrograms, we are interested in mimicking ViV3. Therefore, we used the same perceptual frontend as ViV3. 
%ABISW: need to find the correct word for "time" in "(channels x bands x time)"

%\begin{equation}
%\label{Time_resolution}
%t = \frac{(Sample\_duration - Window\_size)}{Hop\_size} + 1 .
%\end{equation}

During our initial experiments, we observed that in the audio mode of ViV3, the MOS-LQO scores cap at $\approx$4.75 \cite{ViSQOLgithub}. Therefore, we labeled all reference-reference (ref-ref) pairs with the highest MOS score of 5 as ground truth so that our model could rate transparently coded signals to 5. We also observed that both our model and ViV3 were underestimating the quality of very easy-to-code excerpts (e.g. low-frequency dominated content with very low energy at higher frequencies) at very high bitrates. Investigating those excerpts, we observed some visible but inaudible spectral holes (due to coding) in the high-frequency region. To improve the prediction performance of our model, we designed reference-degraded (ref-deg) pairs of signals which are visibly different on Gammatone spectrograms but are perceptually equivalent. Therefore, low-energy (less than $-108$\,dB) high-pass filtered (with $fc=10$\,kHz) noise (of various realizations, e.g., white, pink, and brown) are encoded at high bitrates (with AAC at 80, 96, and 128\, kbps) and added to the training set. After auditioning these ref-deg pairs over headphones and confirming their perceptual equivalence, these ref-deg pairs are labeled manually as MOS score 5. Similarly, digital silence is also included in the training set. The overall composition of the final training set is listed in Table \ref{tbl:training_dataset}.  

\subsection{Test sets}
\label{subsection:testset}
 We used a collection of 56 critical excerpts (typically used to evaluate audio codecs) and the Unified Speech and Audio Coding (USAC) \cite{MPEG-USAC} verification listening tests \cite{usac_lt, USAC} to evaluate the performance of the InSE-NET against ViV3 and subjective listening scores, respectively. These 56 critical excerpts include one applause excerpt, speech excerpts in different languages and genders, music excerpts (including isolated single instruments) as well as excerpts that are a mix of both speech and music. As in the training set, these 56 excerpts were also coded with AAC and HE-AAC for a range of bitrates. This set does not include subjective quality scores. We only used it to evaluate the performance difference between our proposed model and ViV3 on various unseen content-types. 

USAC verification listening tests \cite{usac_lt, USAC} contain 27 excerpts coded with USAC, HE-AAC, and AMR-WB+ with a wide range of bitrates from 8\,kbps mono to 96\,kbps stereo. These comprehensive verification tests were designed to provide information on the subjective performance of the USAC codec. It consists of three separate listening tests: mono at low bitrates and stereo at both low and high bitrates. All tests were MUSHRA tests, whose quality scale ranges from 0 to 100 (where a higher score implies better quality). For the details of these MUSHRA listening tests, interested readers are referred to \cite{usac_lt, USAC}.

\begin{table}[H]
    \setlength\tabcolsep{2pt}
    \caption{Training dataset composition.}
        \vspace{-0.4cm}
        \begin{center}
        \scriptsize{
            \begin{tabular}{|r|c|r|r|}
                \hline
                 \multicolumn{1}{|c|}{\textbf{Pairs}} & \textbf{Example}   & \multicolumn{1}{c|}{\textbf{\begin{tabular}[c]{@{}c@{}}Bitrate\\ {[}kbps{]}\end{tabular}}} & \multicolumn{1}{c|}{\textbf{\begin{tabular}[c]{@{}c@{}}MOS\\ Label\end{tabular}}} \\ \hline\hline
                \begin{tabular}[c]{@{}c@{}}Ref-Ref \\ (music/speech)\end{tabular}  & 5,400 (12\,h)   &   & 5   \\ \hline
                \begin{tabular}[c]{@{}c@{}}Ref-Deg \\ (music/speech)\end{tabular}  & 59,400 (132\, h)   & \begin{tabular}[c]{@{}r@{}}16,20,24\\ 32,40,48\\ 80,96,128\\(+anchors) \end{tabular}   & \begin{tabular}[c]{@{}r@{}}ViV3\\ MOS-LQO\end{tabular}   \\ \hline
                \begin{tabular}[c]{@{}c@{}}Ref-Ref \\ (brown/pink/white\\ noise/silence)\end{tabular}   & 706 (1.58\,h) &   & 5   \\ \hline
                \begin{tabular}[c]{@{}c@{}}Ref-Deg \\ (brown/pink/white\\ noise/silence)\end{tabular}  & 2,118 (4.74\,h)  & 80,96,128   & 5   \\ \hline
            \end{tabular}}
        \end{center}
\label{tbl:training_dataset}
\end{table}

\section{InSE-NET}
\subsection{Model architecture}

\begin{figure}[ht]
  \centering
  \includegraphics[width=.55\linewidth]{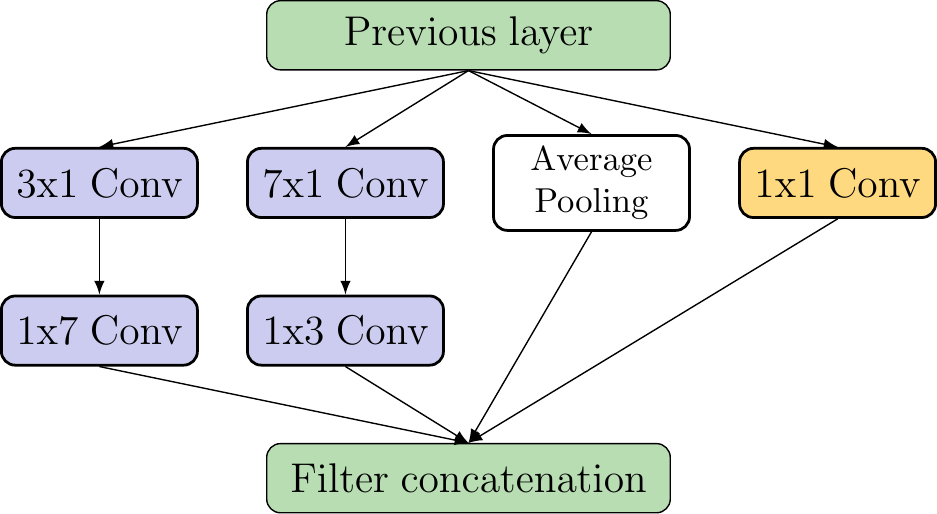} 
  \caption{Modified Inception block with rectangular filters.}
  \label{fig:Inception}
\end{figure}

Our architecture is based on the Inception-v3 \cite{Incept2} network which has been successfully transfer-learned in many audio and video classification tasks such as \cite{Sevilla2017AudioBC} and \cite{HersheyIncept}. During this work, we tried several architectures which have been prevalently used in previous quality evaluation works such as recurrent networks \cite{Fu2018QualityNetAE,NISQA, MittagSiamese} and attention mechanism \cite{MittagSiamese, Roberts2020DeepLS}; but we found the Inception block to be interesting because of its ability to adapt to different receptive field sizes, which is also required for audio. However, we introduced three key changes in our architecture. First, we replaced the square-shaped kernel with vertical \& horizontal rectangular-shaped kernels (3$\times$7, 7$\times$3, 3$\times$5, 5$\times$3). Since audio may contain both transient and tonal signals, exhibited by vertical and horizontal lines on (Gammatone) spectrograms, respectively, we introduced this change to apply both horizontal and vertical kernels for modeling audio. A rectangular kernel such as a horizontal kernel (temporal kernel) could seize rhythmic and tempo patterns and a vertical kernel (frequency kernel) could learn pitch, timbre and such setups over broader frequency bands \cite{RectFilters}. We introduced these rectangular kernels into the structure of the Inception block and developed a new backbone for audio quality evaluation (as depicted in Figure \ref{fig:Inception}). Second, we removed the head layer from the original Inception network. The head layer in the original network was designed to extract features from images. In our case, the Gammatone spectrogram can be already viewed as a feature representation for audio. We found that removing the head layer improved the match with ViV3. Third, we replaced the max-pooling operation with average pooling. The reason being, average pooling retains subtle information in a block. Finally, we split the kernel into smaller ones to reduce the number of parameters (e.g., 3$\times$7 kernel as 3$\times$1 and 1$\times$7).

\begin{figure}[ht]
  \centering
  \includegraphics[width=.35\linewidth]{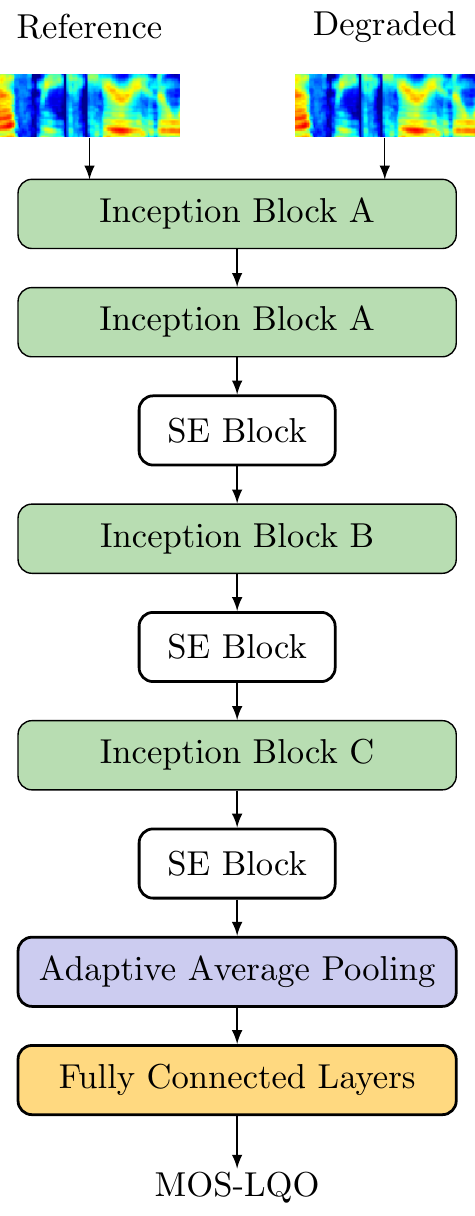} 
  \caption{Block diagram of the InSE-NET model. Input to the model is a pair of Gammatone spectrograms which is computed in a manner identical to ViSQOL v3 \cite{ViSQOLgithub}.}
  \label{fig:InceptionSE}
\end{figure}

The Squeeze-and-excitation (SE) mechanism aims to boost meaningful features while suppressing weak ones by extra attention weights applied to feature maps (channels) \cite{SE}. We inserted SE modules between Inception blocks with minimal complexity overhead (as shown in Figure \ref{fig:InceptionSE}) and report the performance of our proposed InSE-NET. Our proposed model InSE-NET has 15.25M parameters and its architectural and parameter details are tabulated in Table \ref{tbl:parameters}. The kernel sizes employed in the Inception blocks (A, B, and C) considered several possible kernel shapes and sizes (3$\times$3, 3$\times$5, up to 9$\times$9). A parametric grid search was performed to identify the optimal kernel sizes for each layer.  In Table \ref{tbl:parameters}, we directly present the optimal kernel sizes of the Inception blocks. SE layers facilitate the model to focus on those generic features and reduce the influence from the random variations at the instance level. Improving the resistance of the model against such outliers is a trade-off between accuracy and robustness. Extra SE layers are observed to sacrifice refining the model with extremely high accuracy on small, specific scenarios, but pursue a relatively high tolerance on a wider range of unseen data.

\subsection{Training configuration}
The model is trained and evaluated under the following criteria: computation efficiency, number of parameters, mean squared error (MSE), Spearman's correlation coefficient ($R_s$), and Pearson's correlation coefficient ($R_p$). The training set including noise and silence results in a total of 67624 pairs (ref-ref and ref-deg pairs). The training set was partitioned randomly into $80\%$ for training and $20\%$ for validation. 5-fold cross-validation is applied to ensure that the model could make full use of limited data. The average $R_p$ and MSE per fold and epoch were calculated to represent the overall performance of the trained model. The mean and variance of the input features were normalized using the estimation from the training set. All experiments are implemented with PyTorch and were trained for 50 epochs on an Nvidia GTX 1080 Ti GPU using the Adam optimizer. We performed greedy searches over the optimal kernel sizes, learning rates, batch sizes and select a learning rate of $4\times10^{-5}$ and a batch size of 32 as training parameters. Batch normalization was applied after all convolutional layers and smooth L1 loss is used as the loss function. Given the training set size, we use dropout as the regularization technique to prevent overfitting.

\begin{table}[!t]
    \setlength\tabcolsep{1.8pt}
    \vspace{-0.4cm}
    \caption{Architectures and parameters of the InSE-NET model.}
    \vspace{-0.4cm}
    \begin{center}
        \scriptsize{
        \begin{tabular}{|l||r|r|r|r|r|}
            \hline
            \multicolumn{1}{|c||}{\textbf{\backslashbox{\textbf{Layer}}{\textbf{Shape}}}}   & \multicolumn{1}{c|}{\textbf{\begin{tabular}[c]{@{}c@{}}Output\end{tabular}}} & \multicolumn{1}{c|}{\textbf{\begin{tabular}[c]{@{}c@{}}Horizontal \\ Conv.\\ (w$\times$h\,/\,ch)\end{tabular}}} & \multicolumn{1}{c|}{\textbf{\begin{tabular}[c]{@{}c@{}}Vertical \\ Conv. \\ (w$\times$h\,/\,ch)\end{tabular}}} & \multicolumn{1}{c|}{\textbf{\begin{tabular}[c]{@{}c@{}}Normal \\ Conv. \\ (w$\times$h\,/\,ch)\end{tabular}}} & \multicolumn{1}{c|}{\textbf{\begin{tabular}[c]{@{}c@{}} Average \\ Pooling  \\ (w$\times$h)  \end{tabular}}} \\ \hline\hline
            \textbf{Input}   & 2$\times$32$\times$360   &   &   &   &   \\ \hline
            \textbf{Inception A}   & 208$\times$16$\times$180   & 3$\times$7 / 64   & 7$\times$3 / 64   & 1$\times$1 / 64   & 5$\times$5  \\ \hline
            \textbf{Inception A}   & 224$\times$16$\times$90   & 3$\times$7 / 64   & 7$\times$3 / 64   & 1$\times$1 / 64   & 5$\times$5   \\ \hline
            \textbf{SE}   & 224$\times$16$\times$90   &   &   &   &   \\ \hline
            \textbf{Inception B}   & 256$\times$16$\times$45   & 3$\times$5 / 64   & 5$\times$3 / 64   & 1$\times$1 / 64   & 5$\times$5   \\ \hline
            \textbf{SE}   & 256$\times$16$\times$45   &   &   &   &   \\ \hline
            \textbf{Inception C}   & 256$\times$14$\times$22   & 3$\times$3 / 64   & 5$\times$5 / 64   & 1$\times$1 / 64   & 3$\times$3   \\ \hline
            \textbf{SE}   & 256$\times$14$\times$22   &   &   &   &   \\ \hline
            \textbf{\begin{tabular}[c]{@{}l@{}}Adaptive\\ AvgPool\end{tabular}} & 256$\times$4$\times$4   &   &   &   &   \\ \hline
            \textbf{FCL 1}   & 3200$\times$1   &   &   &   &   \\ \hline
            \textbf{FCL 2}   & 512$\times$1   &   &   &   &   \\ \hline
            \textbf{FCL 3}   & 1$\times$1   &   &   &   &   \\ \hline
        \end{tabular}
        }
    \end{center}
    \label{tbl:parameters}
\end{table}

% ABISW: We found that training with MSE-loss (i.e. L_2 loss) didn‘t give good results. Smooth L1-loss gave the best results. 

\section{Results and Discussion}
\subsection{Results on critical excerpts for codecs}
As discussed in section \ref{subsection:testset}, this set consists of 56 excerpts. We compute the correlation between the quality predicted by InSE-NET and that of ViV3 on this test set and examine the InSE-NET under two conditions, namely training with noise and silence and without noise and silence. $R_p$ and $R_s$ are calculated in the Table \ref{tbl:result}, including two low-pass filtered anchors (3.5\,kHz and 7.0\,kHz) and the reference. As can be seen, both $R_p$ and $R_s$ have improved after training with specially designed noise and silence; and the proposed InSE-NET has achieved a strong correlation with ViV3 with a correlation coefficient over 0.97 on the unseen data. These results demonstrate that our model can match ViV3 for all content-types.

\begin{table}[H]
\setlength\tabcolsep{5pt}
\caption{Performance of InSE-NET on the set of \mbox{critical} excerpts for audio codecs. The table shows the correlation coefficients ($R_p$ and $R_s$) between the objective (MOS-LQO) scores, predicted by InSE-NET and that of ViV3.}
\vspace{-0.4cm}
\begin{center}
\scriptsize{
\begin{tabular}{|l||r|r|}
\hline
   \backslashbox{\textbf{Model}}{\textbf{Metric}} & \multicolumn{1}{c|}{\textbf{$\mathbf{R_{p}}$}} & \multicolumn{1}{c|}{\textbf{ $\mathbf{R_{s}}$}} \\ \hline\hline
\textbf{InSE-NET (without noise and silence)}    & 0.937   & 0.919    \\ \hline
\textbf{InSE-NET (with noise and silence)} & 0.972   & 0.985    \\ \hline
\end{tabular}
}
\end{center}
\label{tbl:result}
\end{table}

\begin{figure}[ht]
  \centering
  \includegraphics[width=1.0\linewidth]{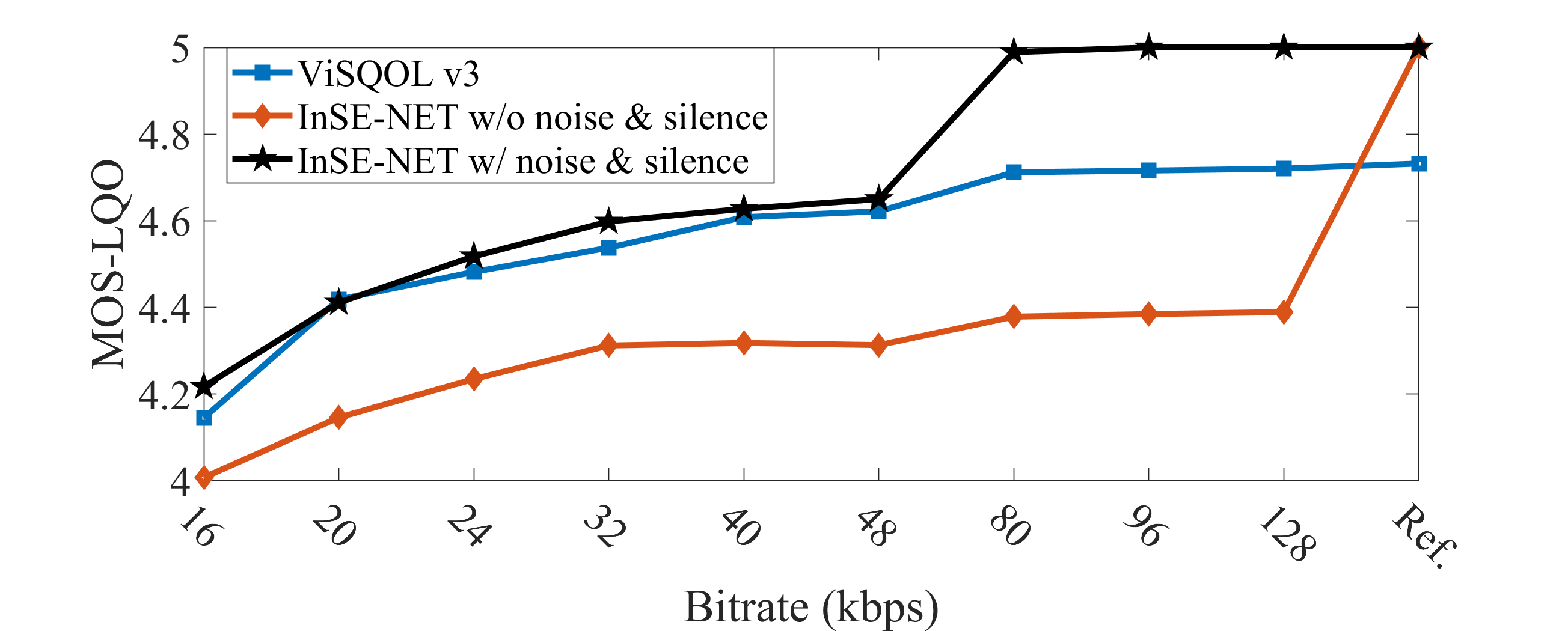} 
  \caption{Prediction of quality of Korean speech, with InSE-NET trained with/without noise and silence. Note that the quality predicted by ViSQOL v3 saturates at 4.732.}
  \label{fig:sc01}
\end{figure}

A Korean speech excerpt is an example where the quality is severely underestimated by our model when trained without noise and silence (see, Figure \ref{fig:sc01}). The rating increases, more obviously (and correctly) for higher bitrates (80, 96, and 128 kbps) after the inclusion of fine-tuned noises and silence in the training set. Recalling from section \ref{subsection:trainingset} (Table \ref{tbl:training_dataset}), we added such synthetic data for these higher bitrates to improve the prediction performance of our model. Thus, Figure \ref{fig:sc01} demonstrates: with carefully designed synthetic data augmentation, one can guide the model to predict a quality score one desires. We observed a similar trend for other low-frequency dominated musical signals. The benefit of this improvement can be observed in the performance of the model on our listening test set.

\subsection{Results on USAC verification listening tests}
In \cite{Fraunhofer}, the authors examined the correlation between different objective scores (PEAQ-Advanced \cite{peaqOnline} and ViA MATLAB \cite{ViSQOLAudioMatlab}) against subjective scores from the mono USAC verification listening test. In Table \ref{tbl:result1}, we list their results together with ViV3 and our InSE-NET. 

\begin{table}[H]
\setlength\tabcolsep{5pt}
\caption{Performance of PEAQ, ViSQOLAudio, ViSQOL-v3, and InSE-NET with respect to the mono USAC verification listening test. The table shows the correlation coefficients ($R_p$ and $R_s$) between predicted objective (MOS-LQO) scores and subjective (MUSHRA) scores from the mono listening test.}
\vspace{-0.4cm}
\begin{center}
\scriptsize{
\begin{tabular}{|l||r|r|}
\hline
   \backslashbox{\textbf{Model}}{\textbf{Metric}} & \multicolumn{1}{c|}{\textbf{$\mathbf{R_{p}}$}} & \multicolumn{1}{c|}{\textbf{$\mathbf{R_{s}}$}} \\ \hline\hline
\textbf{PEAQ Advanced} & 0.650 & 0.700\\  \hline
\textbf{ViSQOLAudio (MATLAB)} & 0.760 & 0.830\\  \hline
\textbf{ViSQOL v3 (C++)} & 0.810 & 0.840\\  \hline
\textbf{InSE-NET} & 0.830 & 0.835\\  \hline
\end{tabular}}
\end{center}
\label{tbl:result1}
\end{table}

Our proposed model has achieved a comparable result as ViV3 and outperformed the older ViSQOLAudio (MATLAB) and PEAQ. Furthermore, we examine the performance of our model and ViV3 on the individual codecs, namely AMR-WB+, HE-AAC, and USAC. Similarly, the reference and two anchors are included in each examination, and corresponding $R_p$ and $R_s$ are listed in Table \ref{tbl:result2}.

\begin{table}[H]
\setlength\tabcolsep{5pt}
\caption{Performance of ViSQOL-v3 and InSE-NET for various codecs regarding the mono USAC verification listening test. The table shows the correlation coefficients ($R_p$ and $R_s$) between predicted objective (MOS-LQO) scores and subjective (MUSHRA) scores from the mono listening test.}
\vspace{-0.4cm}
\begin{center}
\scriptsize{
\begin{tabular}{|l||r|r|r|r|}
\hline
   \backslashbox{\textbf{Codec}}{\textbf{Metric}} & \multicolumn{1}{c|}{\textbf{\begin{tabular}[c]{@{}c@{}}$\mathbf{R_p}$\\ (InSE-NET)\end{tabular}}} & \multicolumn{1}{c|}{\textbf{\begin{tabular}[c]{@{}c@{}}$\mathbf{R_s}$\\ (InSE-NET)\end{tabular}}} & \multicolumn{1}{c|}{\textbf{\begin{tabular}[c]{@{}c@{}}$\mathbf{R_p}$\\ (ViV3)\end{tabular}}} & \multicolumn{1}{c|}{\textbf{\begin{tabular}[c]{@{}c@{}}$\mathbf{R_s}$\\ (ViV3)\end{tabular}}} \\ \hline\hline
\textbf{AMR-WB+}    & 0.889   & 0.856   & 0.877   & 0.862   \\ \hline
\textbf{HE-AAC} & 0.853   & 0.791   & 0.836   & 0.792   \\ \hline
\textbf{USAC}   & 0.873   & 0.881   & 0.853   & 0.881   \\ \hline
\end{tabular}}
\end{center}
\label{tbl:result2}
\end{table}

The proposed model results in a slightly better $R_p$ than ViV3 and has an equivalent $R_s$ as ViV3. Even though we have trained our model with only (HE)-AAC audio codec, the overall performance of our model on different codecs, including AMR-WB+ (a speech codec-based system) is consistent with that of ViV3. Both the proposed model and ViV3 have performed homogeneously well on the experimented codecs with $R_p$ over 0.83 and $R_s$ over 0.79. The estimation of degraded quality by HE-AAC, under this comparison, is unexpectedly the worst among the three experimented codecs, though our model is trained on the excerpts encoded with HE-AAC and AAC. One possible reason lies in that our model takes the ground truth labeled by ViV3 and it captures the exact pattern as to how ViV3 evaluates the quality of signals encoded by HE-AAC. Therefore, the performance of the InSE-NET on different codecs is very similar to that of ViV3, which can be regarded as both a success and currently a limitation of the proposed model. 

\begin{table}[H]
\setlength\tabcolsep{5pt}
\caption{Performance of ViSQOL v3 and InSE-NET on the stereo verification listening tests. The table shows the correlation coefficients ($R_p$ and $R_s$) between predicted objective (MOS-LQO) scores and subjective (MUSHRA) scores from the stereo listening tests.}
\vspace{-0.4cm}
\begin{center}
\scriptsize{
\begin{tabular}{|l||c|r|c|r|}
\hline
\multirow{2}{*}{\backslashbox{\textbf{Model}}{\textbf{Metric}}}   & \multicolumn{2}{c|}{\textbf{Low Bitrates}}   & \multicolumn{2}{c|}{\textbf{High Bitrates}}      \\ \cline{2-5} 
   & $\mathbf{R_p}$   & \multicolumn{1}{c|}{$\mathbf{R_s}$} & $\mathbf{R_p}$   & \multicolumn{1}{c|}{$\mathbf{R_s}$} \\ \hline\hline
\textbf{ViSQOL v3 (C++)} & \multicolumn{1}{r|}{0.777} & 0.782   & \multicolumn{1}{r|}{0.825} & 0.906   \\ \hline
\textbf{InSE-NET}       & \multicolumn{1}{r|}{0.806} & 0.788   & \multicolumn{1}{r|}{0.847} & 0.895   \\ \hline
\end{tabular}}
\end{center}
\label{tbl:result3}
\end{table}

Next, we evaluate the performance on stereo listening tests. Note that our model is neither trained nor designed to specifically handle stereo signals. Internally, ViV3 downmixes stereo (or any multi-channel signal) to a mono mid-signal and then predicts the MOS. So, for stereo, we compute the mid-signal and feed it into our model. InSE-NET results in a slightly better $R_p$ than ViV3 on the two stereo listening tests as listed in Table \ref{tbl:result3}. Moreover, the InSE-NET displays a higher accuracy in estimating the quality of excerpts encoded at high bitrates. Even though our model was neither trained nor designed for stereo, these results display a strong correlation between the prediction of the InSE-NET and the subjective quality score.

Results presented in Tables \ref{tbl:result1}-\ref{tbl:result3}, may give the impression that the proposed InSE-NET model is marginally better than ViV3. However, it is important to note that the proposed data augmentation strategy (see, section \ref{subsection:trainingset}) provides a benefit for low-frequency dominated signals (see, Figure \ref{fig:sc01}). In the USAC verification listening tests \cite{usac_lt, USAC}, only the snippet Siefried02 (a melodic classical music excerpt) is of such nature. For the mono Siefried02 excerpt, the correlation coefficient of the coded audio quality prediction of the InSE-NET model (trained with carefully designed noise and silence) is 0.866, whereas the correlation coefficient of the prediction made by ViV3 is 0.616, thus a 48.58\% improvement in accuracy over ViV3. 

\section{Conclusion}
In this paper, we present InSE-NET, a novel deep network architecture for robustly estimating the perceptual quality of coded audio at a 48\,kHz sample rate for a wide range of bitrates, codecs, and signal categories. Our approach relied on successfully mimicking the MOS of ViV3 as a proxy, followed by improving over ViV3 by augmenting the dataset with carefully designed synthetic data. Our results demonstrate that domain-specific know-how can be incorporated into the dataset design, and more importantly, it can improve the accuracy of the model. At this stage, one may wonder if retraining ViV3 with our proposed data augmentation strategy would have also improved its coded audio quality prediction accuracy. We reiterate here that ViV3 uses Support Vector Regression (SVR) with the maximum range at $\approx$4.75 \cite{ViSQOLgithub}. Thus, simply retraining the SVR model will not completely solve all the weaknesses (e.g., see Figure \ref{fig:sc01}) of ViV3 discussed in the paper. 

We also show the utility of our model for estimating the quality of stereo coding, even though the model was only designed for and trained with mono signals. In the future, we would like to extend our model to stereo and multi-channel coded audio quality prediction and also investigate fine-tuning our model on subjective listening test data. 

Finally, our model enables faster prediction.  An un-optimized PyTorch implementation of our model (excluding ViV3's C++ Gammatone frontend \cite{ViSQOLgithub}) runs at 3x real-time on a CPU. Including a reimplementation of the ViV3's Gammatone frontend in Python, our complete InSE-NET model runs in 2x real-time on a CPU, which is slightly faster than ViV3. However, it is important to reiterate that ViV3 consists of traditional signal processing-based blocks which are fully implemented in C++, whereas our proposed InSE-NET model is a combined (unoptimized) Python and PyTorch implementation of a deep CNN-based model. Hence, even the slight advantage in computational efficiency over ViV3 is appreciable.
\bibliographystyle{IEEEtran}

\newpage

\bibliography{mybib}

\end{document}